\documentclass[prb,onecolumn,superscriptaddress]{revtex4-2}
\usepackage{amsfonts}
\usepackage{amssymb}
\usepackage{amsmath}
\usepackage{graphicx}
\usepackage{dcolumn}
\usepackage{bm}
\usepackage{units}
\usepackage{soul}
\usepackage{multirow}
\usepackage{CJKutf8}
\usepackage{subfigure}
\usepackage{color}%
\usepackage{url}
\usepackage[colorlinks,linkcolor=blue,anchorcolor=blue,citecolor=blue,urlcolor=blue]{hyperref}
\usepackage{array}
\usepackage{ulem}
\usepackage{setspace}
\begin{document}

\title{Nematic and smectic stripe phases and stripe-SkX transformations}

\author{Hai-Tao Wu}
\affiliation{Department of Physics, The Hong Kong University of Science and Technology, 
Clear Water Bay, Kowloon, Hong Kong, 999077, China}
\affiliation{HKUST Shenzhen Research Institute, Shenzhen, 518057, China}

\author{Xu-Chong Hu}
\affiliation{Department of Physics, The Hong Kong University of Science and Technology, 
Clear Water Bay, Kowloon, Hong Kong, 999077, China}
\affiliation{HKUST Shenzhen Research Institute, Shenzhen, 518057, China}

\author{X. R. Wang}
\email{phxwan@ust.hk}
\affiliation{Department of Physics, The Hong Kong University of Science and Technology, 
Clear Water Bay, Kowloon, Hong Kong, 999077, China}
\affiliation{HKUST Shenzhen Research Institute, Shenzhen, 518057, China}

\begin{abstract}
Based on the findings of skyrmion nature of stripes and the metastability of a 
state of an arbitrary number of skyrmions, precisely controlled manipulation of 
stripes of skyrmion number 1 in pre-designed structures and mutual transformation 
between helical states and skyrmion crystals (SkXs) are demonstrated in chiral 
magnetic films. As a proof of the concept, we show how to use patterned magnetic 
fields and spin-transfer torques (STTs) to generate nematic and smectic stripe 
phases, as well as ``UST" mosaic from three curved stripes. Cutting one stripe 
into many pieces and coalescing several skyrmions into one by various external 
fields are good ways to transform helical states and SkXs from each other. \\
\par
\noindent{\bf Spintronic, Landau–Lifshitz–Gilbert Equation Simulation,  
Stripe Skyrmion, Skyrmion Crystal, Nematic Phase.\\}\
\noindent{\bf75.50.Gg 75.70.Ak  75.60.Ch}
\end{abstract}
\date{\today}
\maketitle

\noindent{\large\textbf{1. Introduction}}\\

Magnetic skyrmions, topologically non-trivial spin textures characterized 
by skyrmion number $Q=\frac{1}{4\rm{\pi}}\int \mathbf{m}\cdot(\partial_x 
\mathbf{m} \times \partial_y\mathbf{m})\,{\rm d}x{\rm d}y$, provide a 
fertile ground for studying fundamental physics such as the topological 
Hall effect that is a phenomenon about how non-collinear spins in 
skyrmion crystals (SkXs) affect electron transport \cite{roadmap,the1,
the2}. Here $\mathbf{m}$ is the unit vector of the magnetization. 
Skyrmions were observed in systems involving Dzyaloshinskii–Moriya 
interaction (DMI) \cite{Yu2010,Muhlbauer2009,Heinze,Romming2013,
Kezsmarki2015,Nayak2017} or geometric frustration \cite{Okubo,Leonov,
Kurumaji2019}. Skyrmions are commonly believed to disk-like objects, 
and three families of circular skyrmions have been identified, namely 
spiral (Bloch type) skyrmions, hedgehog (N\'{e}el type) skyrmions, 
and anti-skyrmions \cite{roadmap,Yu2010,Kezsmarki2015,Nayak2017}.
Recently, it is shown that irregular stripes and maze structures are 
also skyrmions with topological skyrmion number 1 \cite{stripe-skm}.
With this expanded zoo, skyrmions provide a useful platform for studying 
fundamental sciences, other than potential applications in information 
technology. For example, one can study the interplay of topology, shape, 
spin and charge. One can ask how topologically non-trivial textures in 
various forms affect electron transport if a precise control of condensed 
skyrmion states is possible? Creation and control of topologically 
non-trivial stripes in long-term searching \cite{roadmap} nematic and 
smectic phases with pre-designed elongation and orientation are the 
main theme of current study. 

Many efforts have been made in skyrmion generations, manipulations, and 
detections \cite{Flovik2017,Jan2016,Kong2013,Zidong2020,Jonietz2010,
Weiwei2015,Chengjie2017,Yuan2019,White2014,Matsukura2015,Ohno2000,
Ando2016,Dohi2016,Weisheit2007,Maruyama2009,Lebeugle2009,Yang2016,
Heron2011,Sch2011,Chiba2012,Franke2015,the2,Woo2016,Schulz2012,
Gongxin2020,Yuan2016,Romming2013,Sampaio2013,Durrenfeld2017,Li,Jiang,
Du2015,Iwasaki2014,Schutte2014,Xichao2015,Shi-Zeng2015,Garanin2017}. 
Magnetic fields \cite{Flovik2017,Jan2016,Kong2013,Zidong2020,Jonietz2010,
Weiwei2015,Chengjie2017,Yuan2019,White2014}, electric fields \cite{Matsukura2015,
Ohno2000,Ando2016,Dohi2016,Weisheit2007,Maruyama2009,Lebeugle2009,Yang2016,
Heron2011,Sch2011,Chiba2012,Franke2015,Yuan2019}, currents \cite{the2,
Woo2016,Schulz2012,Gongxin2020,Yuan2016,Romming2013,Sampaio2013,Durrenfeld2017}, 
geometric constrains \cite{Li,Jiang,Du2015}, spin waves \cite{Iwasaki2014,
Schutte2014,Xichao2015} and temperature gradient \cite{Kong2013,Zidong2020} 
were used to generate and manipulate skyrmions. There were also 
demonstrations of how to use STM to add and delete a skyrmion in  
skyrmion crystals (SkXs) \cite{Romming2013}. With all the advances made 
to date in skyrmion manipulation, the control of helical states and SkXs 
often relies on the luck and a hunch. It is a formidable task to control 
the shape and morphology of individual stripes and the overall 
arrangement of a group of them. This is why the long-time suspected 
liquid-crystal-like skyrmion phases such as nematic or smectic configurations 
have not been found yet \cite{roadmap}. It is also not clear how to precisely 
control transitions from helical states to SkXs. The lack of the ability in 
stripe control is largely due to our ignorance about the skyrmionic nature of 
stripes and the origin of complicated stripe morphologies that include dendrite-like 
and maze structures. Our recent discovery of the skyrmion nature of stripes and 
sensitivity of stripe morphology to skyrmion number density \cite{stripe-skm,
preprint2} provides new thoughts about stripe-state-control and possible 
control of transformations between helical states to SkXs at nanometer scale.  

In this paper, how to use patterned spin-transfer torques (STTs) to create a 
long-term searching nematic and smectic stripe phase is demonstrated. 
The lengths and orientations of stripes can be controlled by the skyrmion 
density and arrangement using patterned fields and STTs. Each stripe has a 
skyrmion number 1. The smectic stripe phase becomes an SkX when the stripe 
length is order of the stripe width. It is also possible to transform an SkX 
to a smectic or a nematic phase by using field pulses to coalesce skyrmions. 
It is even possible to construct a symbol of ``UST" with three curved 
topologically non-trivial stripes.
\\
\par

\noindent{\large\textbf{2. Model and methods}}\\

We consider a thin chiral magnetic film of thickness $d$ in the $xy-$plane. 
Its magnetic energy reads
\begin{equation}
E=d\iint \{A|\nabla\mathbf{m}|^2 + D[m_z\nabla\cdot\mathbf{m}-(\mathbf{m}
\cdot\nabla)m_z]
+K_{\rm u}(1-{m_z^2})-\mu_0M_{\rm s}(\mathbf{H}+\mathbf{H}_{\rm d})
\cdot\mathbf{m}\} \,{\rm d}S,
\label{Etotal}
\end{equation}
where $A$, $D$, $K_{\rm u}$, $\mu_0$, $M_{\rm s}$, $\mathbf{H}$ and 
$\mathbf{H}_{\rm d}$ are the Heisenberg exchange stiffness, the DMI coefficient, 
the perpendicular magneto-crystalline anisotropy, the vacuum permeability, the 
saturation magnetization, the external magnetic field, and the dipolar field, 
respectively. $E=0$ is chosen for state of $\mathbf{m}=\hat z$. In the analytical 
considerations, the static magnetic interaction for a thin film can be included 
through the effective magnetic anisotropy constant $K=K_{\rm u}-\mu_0M_{\rm s}^2/2$ 
(dipolar interaction is fully included in all of our MuMax3 simulations \cite{MuMax3}). 
This theoretical approximation is good when the film thickness $d$ is much smaller 
than the exchange length \cite{skx-size,Xiansi2018}. 

Magnetization unit vector 
$\mathbf{m}$ is govern by the Landau-Lifshitz-Gilbert (LLG) equation 
\begin{equation}
\frac{\partial \mathbf{m}}{\partial t} =-\gamma\mathbf{m} \times 
\mathbf{H}_{\rm eff} +\alpha \mathbf{m} \times \frac{\partial \mathbf{m}}
{\partial t} + \mathbf{\tau},
\label{llg}
\end{equation}
where $\gamma$ and $\alpha$ are respectively the gyromagnetic ratio and the 
Gilbert damping constant. ${\bf H}_{\rm eff}=\frac{2A}{\mu_0M_{\rm s}} 
\nabla^2\mathbf{m}+\frac{2K_u}{\mu_0M_{\rm s}}m_z\hat z+{\bf H}_{\rm DM}+
{\bf H}+{\bf H}_{\rm d}$ is the effective field including the exchange field, 
the magneto-crystalline anisotropy field, the DMI field ${\bf H}_{\rm DM}$,
the external magnetic field {\bf H}, and the magnetic dipolar field 
${\bf H}_{\rm d}$. $\mathbf{\tau}=a\mathbf{m} \times(\mathbf{m}_{\rm p}
\times\mathbf{m})-\beta a \mathbf{m}\times\mathbf{m}_{\rm p}$ is STTs due to 
spin polarized electric current or spin current of polarization $\mathbf{m}_
{\rm p}$ \cite{SLONCZEWSKI1996}, where $a$ proportional to charge current 
density $J$ describes out-of-plane torque, and $\beta$ is a dimensionless 
parameter characterizing the in-plane torque. In this study, 
$\mathbf{m}_{\rm p}=\hat{\it z}$ is assumed if not stated otherwise. 
Eq.~\ref{llg} is numerically solved by Mumax3 package \cite{MuMax3} for 
various patterned currents and magnetic fields. It should be pointed out that 
MuMax3 \cite{MuMax3} includes properly the demagnetization field and considers
both open or periodic boundary conditions. The results in this work are 
for the open boundary conditions (See the Supporting Information). Mesh size 
in this study is $3\,{\rm nm} \times 3\,{\rm nm} \times 1\,{\rm nm}$. 
A large $\alpha=0.3$ is used to speed up stable spin structure search.  
$\alpha$ does not affect the properties of stable and metastable states. 
Skyrmion number of a given spin texture is calculated according to its 
definition, which can be extracted directly from MuMax3 \cite{MuMax3}. 
The material parameters are $A=0.41 \,\rm{pJ}\,\rm m^{-1}$, 
$K_{\rm u}=30\,\rm{kJ}\,\rm m^{-3}$ ($K=4.9\,\rm{kJ}\,\rm m^{-3}$), 
$D=0.12\,\rm{mJ}\,\rm m^{-2}$, and $M_s=0.2\,\rm{MA}\,\rm m^{-1}$ (See 
the Supporting Information), unless otherwise stated.

\bigskip

\begin{figure*}
\includegraphics[width=1\columnwidth]{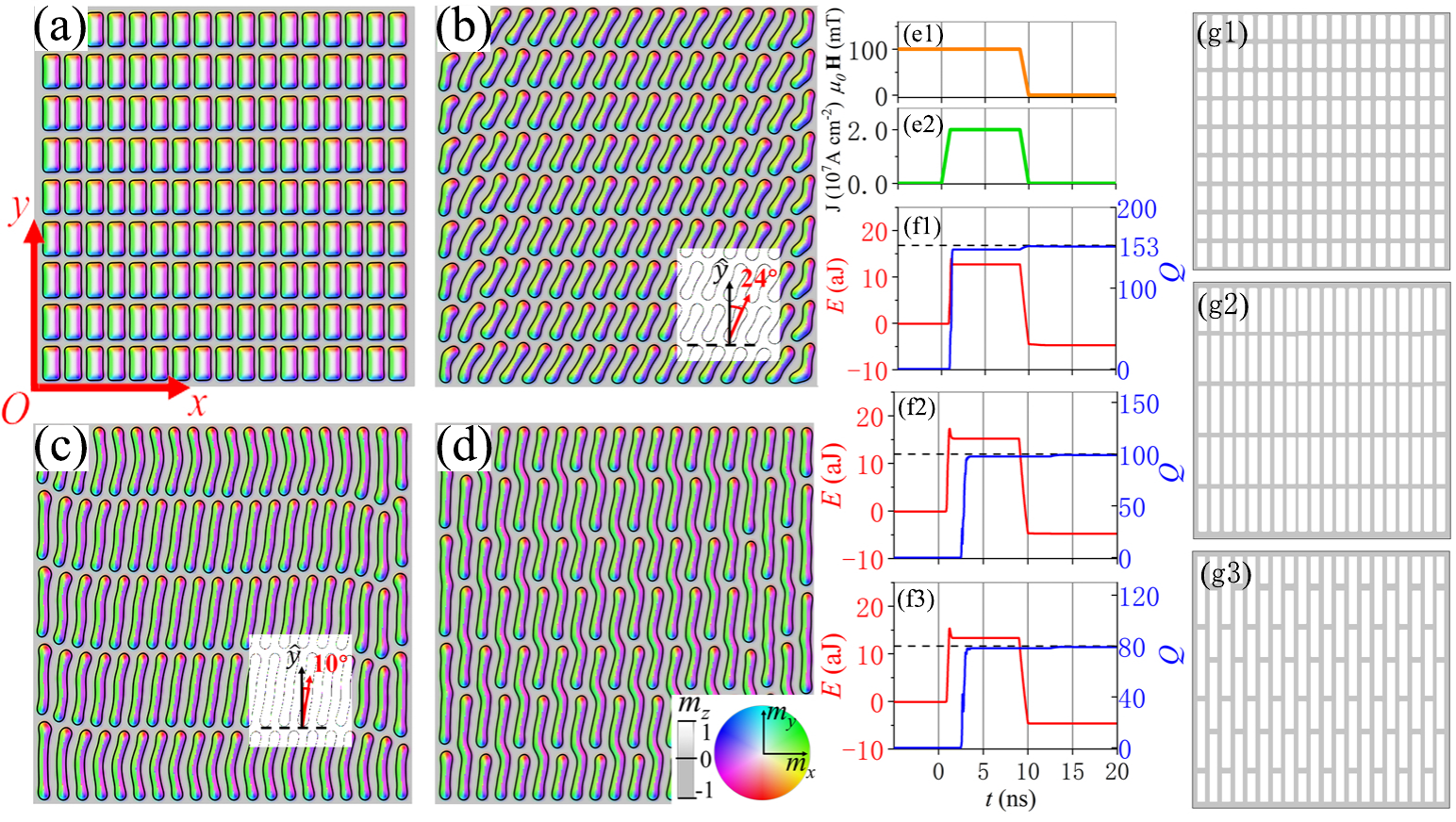}
\caption{\textbf{Creation of smectic and nematic stripe phases.} 
(a) Stripe skyrmions in a stable smectic phase in the presence of a 
magnetic field (e1) and a patterned STT (e2). (b) Stable/metastable 
smectic stripe phase after removal of the magnetic field and the STT. 
Stripes tilt a $24^\circ$ from the $y$-direction. (c) Stable/metastable 
smectic phase of longer stripes after removal of the magnetic field (e1) 
and the patterned STT. Stripes tilt a $10^\circ$ from the $y$-direction.
(d) Stable/metastable nematic phase after removal of the magnetic field 
(e1) and the patterned STT (e2). Color encodes in-plane magnetization 
for $m_z>0$, the background of $m_z<0$ is in gray. Time dependences of 
the magnetic field pulse (e1) and the STT pulse (e2). Evolution of 
total energy $E$ and topological skyrmion number $Q$ for 153 (f1) and 
100 (f2) ordered nucleation domains, and 80 odd-even columns aligned 
nucleation domains (f3). (g1)-(g3) show patterned STTs that produce 
spin textures (b), (c) and (d), respectively. STT torque of 
$a=3.45\times10^{10}$s$^{-1}$ is applied in the white areas.}
\label{Fig1}
\end{figure*}

\noindent{\large\textbf{3. Principles of stripe-state-control}}\\

One of the breakthroughs in skyrmion physics is the understanding of skyrmion 
nature of all kinds of stripes, long and short, curved and straight, ramified and 
non-ramified, in helical, spiral, conical states and in dendrite-like and maze 
structures. Stripes are the natural form of skyrmions when $\kappa\equiv{\pi}^2
D^2/(16AK)>1$, and the ground state is a condensed skyrmion phase with skyrmions 
occupying the whole chiral magnetic film, in contrast to isolated circular 
skyrmions that are metastable in a film with $\kappa\equiv{\pi}^2D^2/(16AK)<1$ 
\cite{stripe-skm,skx-size}. The film with an arbitrary number of skyrmions 
is metastable \cite{stripe-skm,preprint2}. This is very similar 
to a BEC condensate with many atoms staying together (in one energy state).
These stripes, attempting to fill up the whole system, have a well-defined 
width of $L=f(\kappa)A/D$ with $f(\kappa)\approx 2\pi$ for $\kappa\gg1$ 
\cite{stripe-skm,preprint2}. These new understandings are supported by the 
observation of topological charges in stripe structures \cite{Cortesortuno2019,
Jenaeabc2020} and continuous deformation between circular skyrmions and stripes 
\cite{Jena2020,Peng2018,Karubeeaar7043,Yu8856}. No fundamental difference 
between these skyrmions of different shapes is a putative conclusion of the new 
understanding, and transformation from one into another is therefore highly 
possible if a proper kinetic path is used. The final stripe morphology depends on 
skyrmion number density, the initial configuration, and the actual spin dynamics. 
These multiple effects explain the rich morphologies and structures of the 
condensed skyrmion phases. This new understanding of the condensed skyrmion 
phases provides a new methodology of creating and manipulating ordered stripe 
structures by using proper external forces, such as patterned fields and spin 
torques, to control the initial skyrmion seeds and their locations, as well as 
spin dynamics. 
\\
\par

\noindent{\large\textbf{4. Arbitrary skyrmion structures}}\\

As a proof of the concept, a $900\,{\rm nm}\times 900\,{\rm nm} \times 1\,{\rm nm}$ 
chiral magnetic film with parameters given in Model and Methods is considered. 
This film has $\kappa>1$ such that its ground state is stripe skyrmions 
\cite{stripe-skm,preprint2}. It is convenient to use a coordinate system with the 
origin at the left bottom corner such that the film is in $0\,$nm$\le x\le 900\,$nm, 
$0\,$nm$\le y\le 900\,$nm. In this study, the initial state is a perpendicular 
ferromagnetic state of $m_z=-1$ with the help of a strong magnetic field of 
$\mu_0\mathbf{H}=-100\hat{z}\,{\rm mT}$.

At $t=0$, a patterned STT pulse of $10\,$ns long is used to create different 
ordered nucleation domains of $m_z=1$ in the ferromagnetic background. The STT 
increases linearly from 0 to its full value of $a=3.45\times10^{10}\,$s$^{-1}$, 
corresponding to an electric current density of $2.0\times 10^{7}\,$A$\,$cm$^{-2}$, 
in $1\,$ns and linearly decreases later to zero from $t=9\,$ns to $t=10\,$ns 
as shown in Fig.~\ref{Fig1}(e2). $\mu_0\mathbf{H}$ is linearly switched off from 
$t=9\,$ns to $t=10\,$ns as shown in Fig.~\ref{Fig1}(e1). Figure~\ref{Fig1}(a) shows 
153 stripes of $30\,$nm wide and $78\,$nm long in a stable/metastable 
smectic phase at $t=5\,$ns after the STT pulse is applied to 153 rectangular 
areas of $(nl_x-30\,{\rm nm})\le x\le nl_x,~(ml_y-83\,{\rm nm})\le y\le ml_y,
~n=1,\cdots,17 \ \ ~m=1,2,\cdots,9$ with $l_x=52\,$nm and $l_y=98\,$nm. 
The corresponding STT pattern is shown by white rectangulars in 
Fig.~\ref{Fig1}(g1), where an STT torque of $a=3.45\times10^{10}$s$^{-1}$ is applied.
The STT pulse reverses some spins in these areas that evolve into well-aligned 
stripes. The process is captured by the evolutions of system energy 
$E$ and skyrmion number $Q$ as shown in Fig.~\ref{Fig1}(f1). Under the STT 
and $\mu_0\mathbf{H}$, the system quickly reaches the stable smectic phase 
represented by constant $Q$ and $E$. Figure~1(b) is the stable smectic stripe 
phase at $t=15\,$ns in the absence of the magnetic field and the STT. 
Stripes tilt an angle of $24^{\circ}$ from the $\hat y-$axis (the Inset). 
The whole process from Fig.~\ref{Fig1}(a) to Fig.~\ref{Fig1}(b) is captured in 
the first part of Video 1. A smectic phase can 
also be obtained in a bulk sample by the same method where 
each stripe is replaced by a stripe tube that can be viewed as stacked stripes, 
similar to the skyrmion strings in the literature (see the Supporting Information). 

\begin{figure}
\includegraphics[width=\columnwidth]{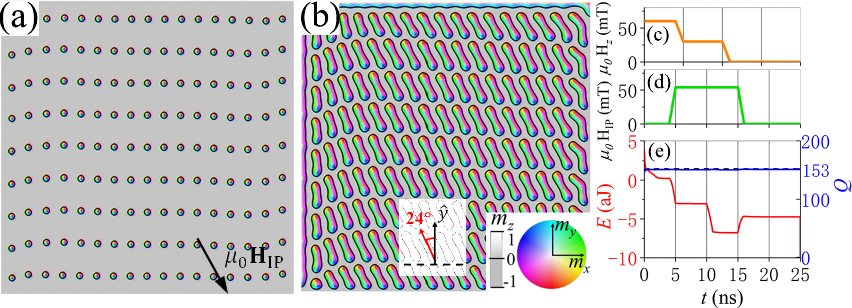}
\caption{\textbf{Transformation between different stripe smectic phases.} 
(a) Circular skyrmions in a crystal structure at $t=4\,$ns after applying a 
field of $-60\hat{z}\,{\rm mT}$ on the smectic phase of Fig. \ref{Fig1}(b). 
(b) The new stable smectic phase at $t=25\,$ns when two series of magnetic 
field pulses of $\mu_0H_z$ (c) and $\mu_0H_{||}$ (d) are removed. 
(e) Evolution of total energy $E$ and skyrmion number $Q$.}
\label{Fig2}
\end{figure}

The length of stripes can be controlled by varying the length of 
rectangles where STT is applied. Figure~\ref{Fig1}(c) is the stable/metastable 
smectic stripe phase in the absence of the field and the STT, after the STT 
pulse Fig.~\ref{Fig1}(e2) is applied on 100 rectangular areas of 
$(nl_x-36\,{\rm nm})\le x\le (nl_x-12\,{\rm nm}),~(ml_y-166\,{\rm nm})\le y\le 
(ml_y-12\,{\rm nm}),~n=1,\cdots,20\ \ ~m=1,2,\cdots,5$ with $l_x=45\,$nm 
and $l_y=180\,$nm. Stripe skyrmions are 162 nm long, longer than those in 
Fig.~\ref{Fig1}(a) and Fig.~\ref{Fig1}(b).
The corresponding STT pattern is shown in Fig.~\ref{Fig1}(g2).
Stripes are inclined away from the vertical direction for very simple physics:
The stripe width $L$ is fixed \cite{stripe-skm,skx-size} while their length is 
stretchable in order to lower their energy due to the negative formation 
energy of skyrmions.  
For $N$ parallel aligned stripes of along $L_x$, stripes will incline away 
from the vertical direction if $L_x>2NL$. The titled angle $\Theta$ satisfies
$2L/(\cos \Theta)=L_x/N$. For stripes in Fig.~\ref{Fig1}(b), $N=17$, 
$L_x=900$nm, $L=23$nm such that $\theta=\cos^{-1}(46/53)\simeq 28^\circ$. 
For stripes in Fig.~\ref{Fig1}(c), $N=20$, $L_x=900$nm, $L=22$nm such that 
$\theta=\cos^{-1}(44/45)\simeq 12^\circ$, which are very close to the observed values.

In order to create a nematic phase with 80 stripes, we apply a field 
pulse of Fig.~\ref{Fig1}(e1) and a patterned STT pulse of Fig.~\ref{Fig1}(e2) 
on 80 rectangular areas: $(nl_x-36\,{\rm nm})\le x\le (nl_x-12\,{\rm nm}), ~18
\,{\rm nm}\le y\le 114{\, \rm nm}~{\rm or}~(ml_y-126\,{\rm nm})\le y\le (ml_y+
111\,{\rm nm})$ for odd $n$, and $792\,{\rm nm}\le y\le 885{\, \rm nm}~{\rm or}
~(ml_y-240\,{\rm nm})\le y\le ml_y$ for even $n$, $n=1,\cdots,20,~m=1,2,3$, 
$l_x=45\,$nm and $l_y=258\,$nm. A stable nematic stripe phase is obtained in 
the absence of the field and the STT as shown in Fig.~\ref{Fig1}(d). 
The corresponding STT pattern is shown in Fig.~\ref{Fig1}(g3).
Skyrmion number $Q$, as well as stripe number, in two smectic phases [Fig.~
\ref{Fig1}(b) and Fig.~\ref{Fig1}(c)] and the nematic phase [Fig.~\ref{Fig1}(d)] 
are 153, 100, and 80, respectively. This supports the claim that all stripes 
have skyrmion number 1. The widths of all stripes in these figures are the 
same about $22\,$nm, smaller than the stripe width in the presence of the
field and the STT that prefers wider skyrmions as shown in Fig.~\ref{Fig1}(a).

It is also possible to transform one stripe phase into another by using external 
forces. For example, starting from the stable smectic phase in Fig.~\ref{Fig1}(b) 
where stripes align $24^\circ$ north-east, stripes shrink into disks as shown in 
Fig.~\ref{Fig2}(a), $4\,$ns after a $\mu_0\mathbf{H}=-60\hat{z}\,{\rm mT}$ field 
pulse shown in Fig.~\ref{Fig2}(c) is applied at $t=0$. Each disk is a circular 
skyrmion of skyrmion number 1 as shown in Fig.~\ref{Fig2}(e) where skyrmion number 
$Q$ and energy $E$ does not change with time. At $t=4\,$ns, $\mu_0H_z$ is linearly 
switched off and an in-plane field pulse of $\mu_0H_{\parallel}=54\,$mT strong 
and $12\,$ns long, pointing $24^\circ$ south-east indicated by the arrow in 
Fig.~\ref{Fig2}(a), is switched on at the same time. The time dependence of two 
field pulses are given in Fig.~\ref{Fig2}(c) and Fig.~\ref{Fig2}(d), respectively. 
A new stable smectic phase where stripes align $24^\circ$ north-west as shown in 
Fig.~\ref{Fig2}(b) is obtained after the removal of external fields. The transformation 
between two smectic phases is recorded in the second part of Video 1.

\begin{figure}
\includegraphics[width=\columnwidth]{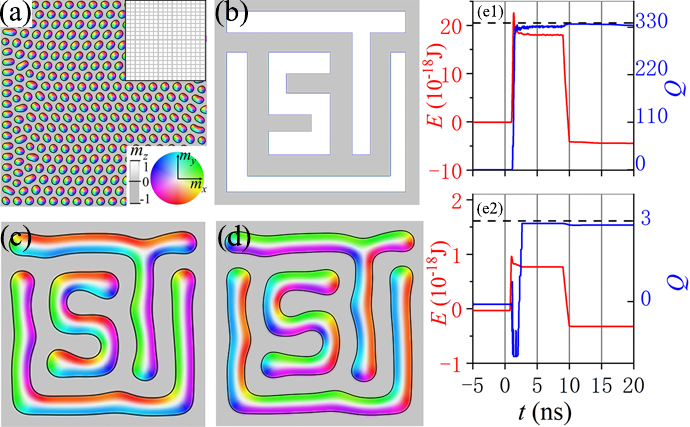}
\caption{ \textbf{Creation of an SkX and a ``UST" mosaic.}
(a) A stable SkX in a $900\,{\rm nm} \times 900\,{\rm nm} \times 1\,{\rm nm}$ 
film at $t=5\,$ns. Inset is the corresponding STT pattern. 
A STT torque pulse of $a=3.45\times10^{10}$s$^{-1}$ is applied in the white areas.
(b) The well-designed mask for producing ``UST" mosaic.
(c) The stable ``UST" mosaic on a $230\,{\rm nm} \times 230\,{\rm nm} \times 1\,
{\rm nm}$ film at $t=20\,$ns after removal of field and STT pulses. 
(d) ``UST" mosaic for the bulk DMI of the same strength as that for (c) at 
$t=20\,$ns after removal of the field and the STT. The only difference between 
(c) and (d) is change of the Neel-type stripe wall to the Bloch-type. 
(e1,e2) Evolution of energy $E$ and skyrmion number $Q$ for SkX and ``UST" 
systems, respectively.}
\label{Fig3}
\end{figure}

Since condensed skyrmions prefer an SkX structure when the skyrmion-skyrmion 
distance is comparable to the stripe width, an SkX can also be obtained from a 
ferromagnetic state of $m_z=-1$ by creating a denser skyrmion nucleation centers
with using the same field and STT pulses described in Fig.~\ref{Fig1}(e1-e2). 
This can be achieved by applying the STT pulses on 340 rectangular areas of 
$(nl_x-30\,{\rm nm})\le x\le nl_x,~(ml_y-42\,{\rm nm})\le y\le (ml_y-12\,{\rm 
nm}),~n=1,2,\cdots,17 ~m=1,2,\cdots,20$ with $l_x=52\,$nm and $l_y=45\,$nm. 
Figure~\ref{Fig3}(a) shows the stable SkX at $t=15\,$ns in the absence of both 
fields and STTs. Inset of Fig.~3(a) shows the corresponding STT pattern.

It is even possible to use stripes to create other more exotic patterns like 
``UST" mosaic, as long as two neighbouring stripes have a distance around 
their natural width. As shown in Fig.~\ref{Fig3}(b), a ``UST" mask is 
designed in a $230\,{\rm nm} \times 230\,{\rm nm}\times 1\,{\rm nm}$ film 
that has the same material parameters as those in Fig.~\ref{Fig1} and \ref{Fig2}. 
The film is initially in the ferromagnetic state of $m_z=-1$ under the field pulse 
shown in Fig.~\ref{Fig1}(e1). At $t=0$, the STT pulse of Fig.~\ref{Fig1}(e2) is applied 
on the mask of Fig.~\ref{Fig3}(b), and Fig.~\ref{Fig3}(c) is the final stable 
pattern $t=15\,$ns after both the magnetic field and the STT are switched off. 
This process is recorded in Video 2. To substantiate our claim that both SkX 
and ``UST" mosaic are stable spin structures, Fig. 3(e1,e2) plot the time 
evolution of energy $E$ (the red and the left axis) and skyrmion number $Q$ 
(the blue and the right axis) for SkX and ``UST" mosaic systems. 
Clearly, both $E$ and $Q$ become constants shortly after $t=10\,$ns. 
Furthermore, it is interesting to notice that ``UST" mosaic is made from 
three stripes, each one of which has skyrmion number 1 and is in shape of 
``U", ``S", and ``T", respectively.

Only interfacial DMI is considered so far, but the results are essentially 
the same in a film with bulk DMI. If the interfacial DMI $D[m_z\nabla\cdot
\mathbf{m}-(\mathbf{m}\cdot\nabla)m_z]$ in Eq.~\ref{Etotal} is replaced by 
the bulk-type DMI of $D[\mathbf{m} \cdot (\nabla \times \mathbf{m})]$, using 
exactly the same mask of Fig.~\ref{Fig3}(b) and the same field and STT pulses 
of Fig.~\ref{Fig1}(e1-e2), similar ``UST" mosaic is obtained as shown in 
Fig.~\ref{Fig3}(d). The only difference is that the Neel-type stripes in the 
interfacial DMI become Bloch-type stripes in a film with bulk DMI 
\cite{stripe-skm}.
\\
\par
\begin{figure}
\includegraphics[width=\columnwidth]{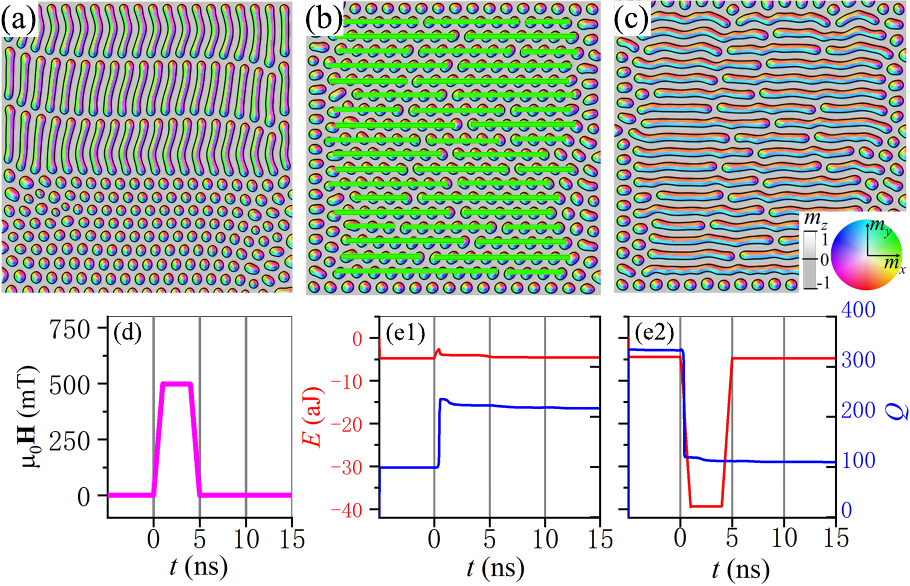}
\caption{\textbf{Transformation between stripe phases and SkXs.} 
(a) The mixture of a smectic stripe phase and an SkX comes from the smectic 
stripe phase in Fig.~\ref{Fig1}(c) by using a field of $-500 \hat z\,$mT to 
cut the two lower rows of stripes into 8 equally spaced shorter pieces. 
(b) The well-designed mask consisting of 42 rectangles of $15\,$nm wide and 
varying length as shown by the green color. A field pulse of $500 \hat z\,$mT 
is applied on the mask to coalesce skyrmions in the SkX. 
(c) The stable nematic stripe phase at $t=15\,$ns after the pulse of (d).
(d) Time dependence of the magnetic field pulse. It linearly rises to full 
value from $t=0$ to $t=1\,$ns, and keep unchanged for $3\,$ns, then decrease 
to zero from $t=4$ to $t=5\,$ns. Evolution of energy and skyrmion number for 
cutting stripes (e1) and coalescing skyrmions (e2). }
\label{Fig4}
\end{figure}

\noindent{\large\textbf{5. Structure transformations}}\\

Cutting stripes into smaller pieces is a good way to transform helical states 
into SkXs. This method is beyond thermodynamic process and can directly drive a 
system from one metastable state into another one. Using the smectic phase of 
Fig.~\ref{Fig1}(c) as an example, a field pulse of $-500\hat{z}\,$mT strong and 
$5\,$ns long is applied in 8 rectangular areas of $0\le x\le 900\,{\rm nm},
~(ml_y-30\,{\rm nm})\le y\le ml_y,~m=1,2,\cdots,8$ with $l_y=42\,$nm. The field 
pulse, whose shape is shown in Fig.~\ref{Fig4}(d), cuts the two lower rows of 
stripes in Fig.~\ref{Fig1}(c) into 8 evenly spacing smaller pieces that become 
circular skyrmions and are in a lattice structure. Figure~\ref{Fig4}(a) shows coexistence 
of a smectic stripe phase and an SkX at $t=15\,$ns. Figure~\ref{Fig4}(e1) shows 
how the energy (the red and the left axis) and the skyrmion number (the blue and the 
right axis) vary with time after the field pulse. 

It is also possible to transfer an SkX into a nematic phase by coalescing 
many skyrmions into one. As exemplified in the SkX of Fig.~\ref{Fig3}(a), a 
strong magnetic field along the $\hat{z}$-direction can coalesce skyrmions. 
Figure~\ref{Fig4}(c) shows the final stable nematic stripe phase after a 
field pulse of the same shape illustrated in Fig.~\ref{Fig4}(d) (but along 
the $\hat{z}$-direction) is applied to a patterned film that consists of 42 
rectangles of $18\,$nm wide and various lengths indicated by the green color in 
Fig.~\ref{Fig4}(b). The field pulse coalesces circular skyrmions into stripe 
skyrmions, and transforms an SkX into a nematic stripe phase. As shown in 
Fig.~\ref{Fig4}(e2), the nematic phase is a stable spin structure of the system 
since both energy $E$ and skyrmion number $Q$ do not vary with time $t=10\,$ns 
after external stimulus is switched off. The film, starting with more than 
300 circular skyrmions before the magnetic field pulse, consists of 42 stripes 
and a number of disk-like skyrmions after the field pulse. 
\\
\par
\noindent{\large\textbf{6. Discussion and conclusion}}\\

All structures we obtained are stable/metastable against thermal agitation.
To substantiate this assertion, we performed MuMax3 simulations at a finite 
temperature \cite{MuMax3,Brown1963}. Video 3 shows the stability of smectic 
phase of Fig.~\ref{Fig1}(b) at $T=5\,$K  ($T_{\rm c}$ is around 
$30\,$K) in the absence of fields and STTs. Stripe skyrmions keep their shape and 
arrangements unchanged under thermal agitation (see the Supporting Information).
Actually, as far as the temperature is not too close to $T_{\rm c}$, 
metastable spin textures keep unchanged for a long time \cite{preprint2}.

Although only magnetic fields and STTs are used here to generate different spin 
structures and to induce transformations from one ordered skyrmion structure into 
another one, other external forces, such as spin-orbit torques, are equally good as 
long as they can induce magnetization reversal such that nucleation domains can 
be created to generate skyrmions. 

\begin{figure*}[t]
\centering
\includegraphics[width=0.4\columnwidth]{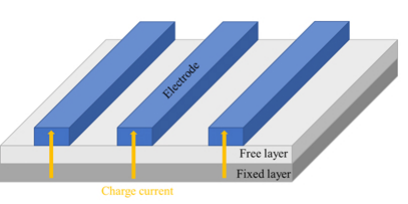}
\caption{\textbf{System illustration: A MTJ with designed 
electrodes.} The lower MTJ is made from a perpendicularly magnetized 
fixed layer at the bottom and a chiral magnetic free layer on the top 
(spacing layer is not shown). Various exotic phases are created in the 
free-layer. A spin polarized charge current after passing through the 
fixed layer can create a patterned STT in the free layer under the electrodes.}
\label{Fig5}
\end{figure*}

One important issue is the feasibility of approaches studied here. 
Firstly, insulating and superconducting masks of nano-meter scale should not 
be a problem for today's technology. In terms of patterned fields or STTs,
one may put masks on both sides of the film using either insulating materials 
or superconducting materials to shield either fields or electric currents such 
that one can realize the desired STT or field patterns. For example, one can 
use standard magnetic tunnelling junction (MTJ) with a ferromagnetic 
fixed layer for STT generation and a chiral magnetic free-layer sandwiching a 
spacing layer. A patterned STT will be generated if a tunnelling current passes 
through the structure. It should not be hard to fabricate structures or masks 
such as the one shown in Fig.~\ref{Fig5} at nanometer scale with state-of-art 
photolithography technologies. Also, masks for generating those patterned STTs 
and fields shown in the Figs.~\ref{Fig1}-\ref{Fig4} should be also an easy task 
with using existing nano-fabrication facilities such as all kinds of lithography 
technologies, ion beam, and all kinds of etching methods. 

Apart from generating the long-term searching nematic/smectic phase, 
one potential application of our result is neuromorphic computing. 
Stripes may work as non-volatile synapse encoding the synapse 
connections by stripe shapes and orientations through the tunnelling 
magnetoresistance or Hall resistance. One attraction of stripe synapses 
is the controllability by stimulus, such as fields and spin torques. 
Stripe-based neuromorphic devices shall be more robust than isolated 
skyrmions-based devices, because of the higher energy barrier 
and entanglements.

In summary, how to use patterned magnetic field and STT pulses to create various 
ordered stripe phases is demonstrated. Based on the stripe nature of skyrmions 
in the ground state of a chiral magnetic film that can host an arbitrary number 
of skyrmions, the creation of long-term searching nematic and smectic stripe phases 
is demonstrated. Cutting long stripes into shorter pieces or coalescing many 
small skyrmions into one stripe skyrmion is a useful way to transform various 
stripe structures from each other and into SkXs, and vice versa. It is also 
demonstrated how to create curved stripes that form a ``UST" mosaic. These 
findings provide a guidance to skyrmion manipulations. 
\par
\bigskip

%

\bigskip
\noindent{\large\textbf{Additional information}}\\
This work is supported by the National Key Research and Development 
Program of China (Grant Nos. 2018YFB0407600 and 2020YFA0309600), 
the NSFC Grant (No. 11974296 and 11774296), and Hong Kong RGC Grants 
(No. 16301518 and 16301619). \\
\textbf{Supplementary Information}\\
The supporting information is available online at http://
\\
\par
\noindent{\textbf{Conflict of Interest:}
The authors declare that they have no 
conflict of interest.
}

\end{document}